\documentclass{ws-procs9x6}

\begin{document}

\title{Leptonic CP Violation and Leptogenesis}

\author{M. N. Rebelo$^*$ }

\address{Departamento de F\' \i sica and Centro de F\' \i sica 
Te\' orica de Part\' \i culas (CFTP)\\
Instituto Superior T\' ecnico (IST), Av. Rovisco Pais, 1049-001 Lisboa, 
Portugal\\
CERN, Department of Physics, Theory Division, CH-1211 Gen\` eve 23, 
Switzerland \\
$^*$Presently at CERN on sabbatical leave from IST.\\
E-mail: margarida.rebelo@cern.ch and rebelo@ist.utl.pt}

\begin{abstract}
We review some recent results on the connection
between CP violation at low energies and Leptogenesis in the
framework of specific flavour structures for the fundamental
leptonic mass matrices with zero textures.
\end{abstract}


\bodymatter

\section{Introduction}\label{aba:sec1}
Neutrinos have masses which are much smaller than the other 
fermionic masses and there is large mixing in the leptonic sector.
The Standard Model (SM) of electroweak interactions cannot 
accommodate the observed neutrino masses and leptonic mixing since
in the Standard Model neutrinos are strictly massless:
the absence of righthanded components for the neutrino
fields does not allow one to write a Dirac mass term; the fact that 
the lefthanded components of the neutrino fields are part of a
doublet of $SU(2)$ rules out the possibility of introducing 
Majorana mass terms since these would violate gauge symmetry;
finally, in the SM, $B-L$ is exactly conserved, therefore 
Majorana mass terms cannot be generated neither radiatively in higher
orders nor nonperturbatively. Therefore,
neutrino masses require physics beyond the SM. At present, this is 
the only direct evidence for physics beyond the SM. The origin of 
neutrino masses remains an open question. It is part of a wider 
puzzle, the flavour puzzle, with questions such as whether or not
there is a connection between quarks and leptons explaining the different
patterns of flavour mixing in each sector and the different mass
hierarchies. In the seesaw framework \cite{Minkowski:1977sc, Yanagida:1979as,
Levy:1980ws, VanNieuwenhuizen:1979hm, Mohapatra:1979ia}
the explanation of the observed smallness of neutrino masses 
is related  to the existence of heavy neutrinos with masses that can be
of the order of the unification scale and have profound implications 
for cosmology. Mixing in the leptonic sector leads to the possibility of 
leptonic CP violation both at low and at high energies. CP violation
in the decay of heavy neutrinos may allow for the  explanation
of the observed baryon asymmetry of the Universe (BAU) through leptogenesis 
\cite{Fukugita:1986hr}.
Neutrino physics may also be relevant to the understanding of dark matter 
and dark energy as well as galaxy-cluster formation.
Recent detailed analyses of the present theoretical and experimental
situation  in neutrino physics and its 
future, can be found in  Refs.~\refcite{Mohapatra:2005wg} and 
\refcite{Group:2007kx}.

In this work the possibility that BAU may be generated via
leptogenesis through the decay of heavy neutrinos is discussed. 
Leptogenesis requires CP 
violation in the decays of heavy neutrinos. However, in general it is not 
possible to establish a connection between CP violation 
required for leptogenesis and low energy CP violation 
\cite{Branco:2001pq, Rebelo:2002wj}. 
This connection
can only be established in specific flavour models. The fact that 
in this framework the masses of the heavy neutrinos are so large that they 
cannot be produced at present colliders and
would have decayed in the early Universe shows the relevance of
flavour models in order to prove leptogenesis. In what follows
it will be shown how the imposition of texture zeros in the neutrino 
Yukawa couplings may at the same time constrain physics at low energies 
and lead to predictions for leptogenesis.

\section{Framework and Notation}
The work described here is done in the seesaw framework, which
provides an elegent way to explain the
smallness of neutrino masses, when compared to the masses of 
the other fermions.

In the minimal seesaw framework, the SM is extended only through the
inclusion of righthanded components for the neutrinos which are 
singlets of $SU(2) \times U(1)$. Frequently,
one righthanded neutrino component per generation is introduced.
This will be the case in what follows, unless otherwise stated.
In fact, neutrino masses can be generated without requiring the
number of righthanded and lefthanded neutrinos to be equal. Present
observations  are consistent with the introdution of two righthanded 
components only. In this case one of the three light neutrinos would be 
massless.

With one righthanded neutrino component per generation the
number of fermionic degrees of freedom for neutrinos
equals those of all other fermions in the theory. However
neutrinos are the only known fermions which have zero
electrical charge and this allows one to write Majorana 
mass terms for the singlet fermion fields.
After spontaneous symmetry breakdown (SSB) the leptonic mass term
is of the form:
\begin{eqnarray}
{\cal L}_m  &=& -[\overline{{\nu}_{L}^0} m_D \nu_{R}^0 +
\frac{1}{2} \nu_{R}^{0T} C M_R \nu_{R}^0+
\overline{l_L^0} m_l l_R^0] + h. c. = \nonumber \\
&=& - [\frac{1}{2}  n_{L}^{T} C {\cal M}^* n_L +
\overline{l_L^0} m_l l_R^0 ] + h. c.
\label{lrd}
\end{eqnarray}
with the $6 \times 6$ matrix $\cal M $  given by:
\begin{eqnarray}
{\cal M}= \left(\begin{array}{cc}  
 0 & m_D \\
m^T_D & M_R \end{array}\right) \label{calm}
\end{eqnarray}
the upperscript $0$ in the neutrino ($\nu$) and charged lepton fields ($l$)
is used to indicate  that we are still in a weak basis (WB), 
i.e., the gauge currents
are still diagonal. The charged current is given by:
\begin{equation}
{\cal L}_W = - \frac{g}{\sqrt{2}}  W^+_{\mu} \ 
 \overline{l^0_L} \ \gamma^{\mu} \ \nu^0_{L} +h.c.
\label{16}
\end{equation} 
Since the Majorana mass term is gauge invariant there are no
constraints on the scale of $M_R$. The seesaw limit consists
of taking this scale to be much larger than the scale of the
Dirac mass matrices $m_D$ and $m_l$. The Dirac mass matrices 
are generated from Yukawa couplings after SSB and are therefore 
at most of the electroweak scale. As a result the
spectrum of the neutrino masses splits into two sets, one consisting of
very heavy neutrinos with masses of the order of that of the matrix $M_R$
and the other set with masses obtained, 
to a very good approximation, from the diagonalisation
of an effective Majorana mass matrix given by:
\begin{equation}
m_{eff} = - m_D \frac{1}{M_R} m^T_D \label{meff}
\end{equation} 
This expression shows that the light neutrino masses are
strongly suppressed with respect to the electroweak scale.
There is no loss of generality in choosing 
a WB where $m_l$ is real diagonal and 
positive. The diagonalization of
${\cal M}$ is performed via the unitary transformation:
 \begin{equation}
V^T {\cal M}^* V = \cal D \label{dgm}
\end{equation}
where ${\cal D} ={\rm diag} (m_1, m_2, m_3,
M_1, M_2, M_3)$,
with $m_i$ and $M_i$ denoting the physical
masses of the light and heavy Majorana neutrinos, respectively. It is
convenient to write $V$ and $\cal D$ in the following block form:
\begin{eqnarray}
V=\left(\begin{array}{cc}
K & G \\
S & T \end{array}\right) ; \qquad 
{\cal D}=\left(\begin{array}{cc}
d & 0 \\
0 & D \end{array}\right) . \label{matd}
\end{eqnarray}
The neutrino weak-eigenstates are then related
to the mass eigenstates by:
\begin{eqnarray}
{\nu^0_i}_L= V_{i \alpha} {\nu_{\alpha}}_L=(K, G)
\left(\begin{array}{c}
{\nu_i}_L  \\
{N_i}_L \end{array} \right) \quad \left(\begin{array}{c} i=1,2,3 \\
\alpha=1,2,...6 \end{array} \right)
\label{15}
\end{eqnarray}
and the leptonic charged current interactions are given by:
\begin{equation}
{\cal L}_W = - \frac{g}{\sqrt{2}} \left( \overline{l_{iL}} 
\gamma_{\mu} K_{ij} {\nu_j}_L +
\overline{l_{iL}} \gamma_{\mu} G_{ij} {N_j}_L \right) W^{\mu}+h.c.
\label{phys}
\end{equation}
with $K$ and $G$ being the charged current couplings of charged 
leptons to the light neutrinos $\nu_j$ and to the heavy neutrinos 
$N_j$, respectively.

In the seesaw limit the matrix $K$ coincides to an excellent approximation 
with the unitary matrix $U_{\nu}$
that diagonalises $m_{eff}$ of Eq.~(\ref{meff}):
\begin{equation} 
-U_{\nu}^\dagger \ m_D \frac{1}{M_R} m^T_D \  U_{\nu}^* =d \label{14}
\end{equation} 
and the matrix $G$ verifies the exact relation:
\begin{equation}
G=m_D T^* D^{-1} \label{exa}
\end{equation} 
and is therefore very suppressed.

In a general framework, with 
${\cal M}$ symmetric, without the zero block present in Eq.~(\ref{calm})
the $3 \times 6$ physical matrix $(K, G)$ of the $6 \times 6$ unitary 
matrix $V$ would depend on six independent mixing angles and
twelve independent CP violating phases  \cite{Branco:1986my}.
This would be possible with a further extention of the SM
including a Higgs triplet. The presence of the zero block reduces
the number of independent CP violating phases to six  \cite{Endoh:2000hc}.
In the seesaw framework massive neutrinos lead to
the possibility of CP violation in the leptonic sector both at low and at
high energies. CP violation at high energies manifests itself in the
decays of heavy neutrinos and is sensitive to phases appearing
in the matrix $G$.

\section{Low Energy Leptonic Physics}
The light  neutrino masses are obtained from the diagonalisation 
of $m_{eff}$ defined by Eq.~(\ref{meff}) which is an effective Majorana
mass matrix. The unitary matrix $U_{\nu}$ that diagonalises  $m_{eff}$
in the WB where the charged lepton masses are already 
diagonal real and positive is known as the
Pontecorvo, Maki, Nakagawa, Sakata (PMNS) matrix \cite{pmns},
and can be parametrised as \cite{Yao:2006px}:
\begin{small}
\begin{eqnarray}
U_{\nu} =\left(
\begin{array}{ccc}
c_{12} c_{13} & s_{12} c_{13} & s_{13} e^{-i \delta}  \\
-s_{12} c_{23} - c_{12} s_{23} s_{13}   e^{i \delta}
& \quad c_{12} c_{23}  - s_{12} s_{23}  s_{13} e^{i \delta} \quad 
& s_{23} c_{13}  \\
s_{12} s_{23} - c_{12} c_{23} s_{13} e^{i \delta}
& -c_{12} s_{23} - s_{12} c_{23} s_{13} e^{i \delta}
& c_{23} c_{13} 
\end{array}\right) \ \cdot \ P
\label{std}
\end{eqnarray}
\end{small}
with $P = \mathrm{diag} \ (1,e^{i\alpha}, e^{i\beta})$,
$\alpha $ and $\beta$ are phases associated to the
Majorana character of neutrinos. There are three CP violating phases
in  $U_{\nu}$.

Experimentally it is not yet known whether any of the three CP violating 
phases of the leptonic sector is different from zero.
The current experimental bounds on neutrino masses and leptonic
mixing are \cite{Yao:2006px}:
\begin{eqnarray}
\Delta m^2_{21} & = & 8.0 ^{+0.4}_{-0.3} \times 10^{-5}\  {\rm eV}^2 \\
\sin^2 (2 \theta_{12}) & = & 0.86 ^{+0.03}_{-0.04} \\ 
|\Delta m^2_{32}| & = & (1.9 \ \  \mbox{to} \ \  3.0) \times 10^{-3}\  
{\rm eV}^2 \\
\sin ^2 ( 2 \theta_{23}) & > & 0.92 \\ 
\sin ^2  \theta_{13} &  < & 0.05 
\end{eqnarray} 
with $\Delta m^2_{ij} \equiv m^2_j - m^2_i$. The angle $ \theta_{23} $ 
may be maximal, meaning $45^{\circ}$, whilst  
$ \theta_{12} $ is already known to deviate from this value. At the moment, 
there is only an experimental upper bound on the angle $ \theta_{13}$.

It is also not yet known
whether the ordering of the light neutrino masses is normal, i.e,
$m_1<m_2<m_3$ or inverted $m_3<m_1<m_2$. The scale of the neutrino
masses is also not yet established. Direct kinematical limits from
Mainz  \cite{Kraus:2004zw} and Troitsk \cite{Lobashev:1999tp}
place an upper bound on $m_{\beta}$ defined as:
\begin{equation}
m_{\beta} \equiv \sqrt{\sum_{i} |U_{ei}|^2 m^2_i}
\end{equation}
given by $m_{\beta} \leq 2.3$ eV (Mainz), 
$m_{\beta} \leq 2.2$ eV  (Troitsk). The forthcoming KATRIN experiment
\cite{Osipowicz:2001sq} is expected to be sensitive to
$m_{\beta} > 0.2$ eV and to start taking data in 2010 \cite{Valerius:2007fw}. 

It is possible to obtain information on the absolute scale of
neutrino masses from the study of the cosmic microwave radiation 
spectrum together with the study of the large scale structure of the
universe. For a flat universe, WMAP combined with other astronomical 
data leads to \cite{Spergel:2006hy} $\sum_{i} m_i \leq 0.66 $ eV
($95\%$ CL). 

Neutrinoless double beta decay can also provide
information on the absolute scale  of the neutrino masses.
In the present framework, in the absence of additional lepton number
violating interactions, it provides a measurement of the
effective Majorana mass given by:
\begin{equation}
m_{ee} = \left| m_1 U_{e1}^2 +  m_2 U_{e2}^2  +  m_3 U_{e3}^2 \right|
\end{equation}
The present upper limit is $m_{ee} \leq 0.9$ eV
\cite{Fogli:2004as} from the Heidelberg-Moskow 
\cite{KlapdorKleingrothaus:2000sn} and the IGEX \cite{Aalseth:2002rf}
experiments. There is a claim of discovery of neutrinoless
double beta decay by the Heidelberg-Moscow collaboration
\cite{KlapdorKleingrothaus:2004wj}. Interpreted in terms of
a Majorana mass of the neutrino, this implies $m_{ee}$
between 0.12 eV to 0.90 eV. This result awaits confirmation
from other experiments and would constitute a major discovery.

It was shown that the strength of CP violation at low energies,
observable for example through neutrino oscillations can be obtained
from the following low energy WB invariant \cite{Branco:1986gr}:
\begin{equation}
Tr[h_{eff}, h_l]^3= - 6i \Delta_{21} \Delta_{32} \Delta_{31}
{\rm Im} \{ (h_{eff})_{12}(h_{eff})_{23}(h_{eff})_{31} \} \label{trc}
\end{equation}
where $h_{eff}=m_{eff}{m_{eff}}^{\dagger} $, 
$ h_l = m_l m^\dagger_l $, and
$\Delta_{21}=({m_{\mu}}^2-{m_e}^2)$ with analogous expressions for
$\Delta_{31}$, $\Delta_{32}$. The righthand side of this equation
is the computation of this invariant in the special WB where the 
charged lepton masses are real and diagonal. 
In the case of no CP violation of Dirac type
in the leptonic sector this WB invariant vanishes; on the 
other hand, it is not sensitive to the presence of 
Majorana phases. This quantity can be 
computed in any WB and therefore is extremely useful for model
building since it enables one to investigate whether a specific
ansatz leads to Dirac type CP violation or not, without the need
to go to the physical basis. 
It is also possible to write WB invariant conditions 
sensitive to the Majorana phases. The general procedure was 
outlined in Ref.~\refcite{Bernabeu:1986fc} where it was applied to the 
quark sector.  For three generations it
was shown that the following four conditions are sufficient
\cite{Branco:1986gr} to guarantee CP invariance:
\begin{eqnarray}
{\rm Im \ tr } \left[ h_l \; (m_{eff} \; m^*_{eff}) \;
( m_{eff} \; h^*_l \; m^*_{eff})\right] & = & 0 \label{41} \\
{\rm Im \ tr } \left[ h_l \; (m_{eff} \; m^*_{eff})^2 \;   
( m_{eff} \; h^*_l \; m^*_{eff}) \right] & = & 0 \label{42} \\
{\rm Im \ tr } \left[ h_l \; (m_{eff} \; m^*_{eff})^2 \   
( m_{eff} \; h^*_l \; m^*_{eff}) \; (m_{eff} \; m^*_{eff})\right]
 & = & 0 \label{43} \\
{\rm Im \ det } \left[ ( m^*_{eff} \; h_l \; m_{eff}) 
+ (h^*_l \;  m^*_{eff} \; m_{eff} )\right]   & = & 0 \label{44} 
\end{eqnarray}
provided that neutrino masses are nonzero and nondegenerate (see also
Ref.~\refcite{Dreiner:2007yz}). In Ref.~\refcite{Branco:2004hu}
alternative WB invariant conditions necessary to guarantee
CP invariance in the leptonic sector under less general circumstances
are given.

\section{Leptogenesis}
The observed baryon asymmetry of the universe (BAU) is given by
\cite{Bennett:2003bz}:
\begin{equation}
\frac{n_{B}-n_{\overline B}}{n_{\gamma}}= (6.1 ^{+0.3}_{-0.2}) \times 10^{-10}.
\end{equation}   
It is already established that this observation requires 
physics beyond the SM in order to be explained. 
One of the most plausibe explanations is 
Leptogenesis \cite{Fukugita:1986hr} where out-of-equilibrium 
L-violating decays of heavy Majorana neutrinos 
generate  a lepton asymmetry which is partially converted 
through sphaleron processes \cite{Kuzmin:1985mm}
into a baryon asymmetry.
The lepton number asymmetry $\varepsilon _{N_{j}}$,
thus produced was computed by
several authors \cite{Liu:1993tg, 
Flanz:1994yx, Covi:1996wh, Pilaftsis:1997jf, Buchmuller:1997yu}.
Summing over all charged leptons
one obtains for the asymmetry produced by the decay of the
heavy Majorana neutrino $N_j$ into the charged leptons 
$l_i^\pm$ ($i$ = e, $\mu$, $\tau$):
\begin{small}
\begin{eqnarray}
\varepsilon _{N_{j}} = & \nonumber \\
= \frac{g^2} {{M_W}^2} & \sum_{k \ne j} \left[
{\rm Im} \left((m_D^\dagger m_D)_{jk} (m_D^\dagger m_D)_{jk} \right)
\frac{1}{16 \pi} \left(I(x_k)+ \frac{\sqrt{x_k}}{1-x_k} \right)
\right]
\frac{1}{(m_D^\dagger m_D)_{jj}} =  \nonumber \\
= \frac{g^2}{{M_W}^2} & \sum_{k \ne j} \left[ (M_k)^2
{\rm Im} \left((G^\dagger G)_{jk} (G^\dagger G)_{jk} \right)
\frac{1}{16 \pi} \left(I(x_k)+ \frac{\sqrt{x_k}}{1-x_k} \right)
\right]
\frac{1}{(G^\dagger G)_{jj}} \nonumber \\ 
\label{rmy}
\end{eqnarray}
\end{small}
where $M_k$ denote the heavy neutrino masses,
the variable $x_k$
is defined as  $x_k=\frac{{M_k}^2}{{M_j}^2}$ and
$ I(x_k)=\sqrt{x_k} \left(1+(1+x_k) \log(\frac{x_k}{1+x_k}) \right)$.
From Equation (\ref{rmy})
it can be seen that, when one sums over all 
charged leptons, the lepton-number
asymmetry is only sensitive to the CP-violating phases
appearing in $m_D^\dagger m_D$ in the WB, where $M_R $
is diagonal.
Weak basis invariants relevant for leptogenesis  
were derived in \cite{Branco:2001pq}:
\begin{eqnarray}
I_1 \equiv {\rm Im Tr}[h_D H_R M^*_R h^*_D M_R]=0  \\
I_2 \equiv {\rm Im Tr}[h_D H^2_R M^*_R h^*_D M_R] = 0 \\
I_3 \equiv {\rm Im Tr}[h_D H^2_R M^*_R h^*_D M_R H_R] = 0 
\end{eqnarray}
with $h_D = m^\dagger_D m_D$ and  $H_R = M^\dagger_R M_R$.
These constitute a set of necessary and sufficient conditions
in the case of three heavy neutrinos. See also \cite{Pilaftsis:1997jf}.

The simplest realisation of thermal leptogenesis consists of
having hierarchical heavy neutrinos. In this case there is 
a lower bound for the mass of the lightest of the heavy neutrinos
\cite{Davidson:2002qv, Hamaguchi:2001gw}. Depending on the
cosmological scenario, the range for minimal $M_1$ varies from
order $10^7$ Gev to $10^9$ Gev \cite{Buchmuller:2002rq, Giudice:2003jh}.
Furthermore, an upper bound on the light neutrino masses is obtained in
order for leptogenesis to be viable. With the assumption that
washout effects are not sensitive to the different flavours
of charged leptons into which the heavy neutrino decays
this bound is approximately $0.1$ ev
\cite{Buchmuller:2003gz, Hambye:2003rt,
Buchmuller:2004nz, Buchmuller:2004tu}. However,
it was recently pointed out  \cite{Barbieri:1999ma, Endoh:2003mz,
Fujihara:2005pv, Pilaftsis:2005rv, Vives:2005ra,
Abada:2006fw, Nardi:2006fx, Abada:2006ea, Blanchet:2006be}
that there are cases where flavour matters and the commonly
used expressions for the lepton asymmetry, which depend on the
total CP asymmetry and one single efficiency factor, may
fail to reproduce the correct lepton asymmetry. In this cases,
the calculation of the baryon asymmetry with hierarchical
righthanded neutrinos must take into consideration flavour
dependent washout processes. As a result, in this case,
the previous upper limit on the light neutrino masses does
not survive and leptogenesis can be made  viable with neutrino masses
reaching the cosmological bound of $\sum_{i} m_i \leq 0.66 $ eV.
The lower bound on $M_1$ does not move much with the inclusion of 
flavour effects.  Flavour effects bring new sources 
of CP violation to leptogenesis and the possibility of having a 
common origin for CP violation at low energies and 
for leptogenesis \cite{Pascoli:2006ie, Branco:2006hz,
Branco:2006ce, Uhlig:2006xf}.

There are very interesting alternative scenarios to the minimal
leptogenesis scenario briefly mentioned here. It was pointed out  
at this conference that an SU(2)-singlet neutrino with a keV
mass is a viable dark matter candidate \cite{Kusenko:2007ay}.
Some leptogenesis scenarios are compatible with much lower 
heavy neutrino masses than the values required for minimal 
leptogenesis.

\section{Implications from Zero neutrino Yukawa Textures}
The general seesaw framework contains a large number
of free parameters. The introduction of zero textures
and/or the reduction of the number of righthanded neutrinos to two,
allows to reduce the number of parameters. In this work only zero 
textures imposed in the fundamental leptonic mass matrices
are considered and, in particular, zero textures of the Dirac neutrino mass
matrix, $m_D$ in the WB where $M_R$ and $m_l$ are real and diagonal.
Zero textures of the low energy effective neutrino mass matrix
are also very interesting phenomenologically \cite{Frampton:2002yf}.
The physical meaning of the zero textures that appear in most
of the leptonic mass ans\" atze was analysed in a recent work  
\cite{Branco:2007nn} where it is shown that some leptonic 
zero texture  ans\" atze can be 
obtained from WB  transformations and therefore 
have no physical meaning.

In general, zero textures reduce the number of  CP violating 
phases, as a result some sets of zero textures imply the vanishing
of certain  CP-odd WB invariants  \cite{Branco:2005jr}. 
This is an important fact since
clearly zero textures are not WB invariant, therefore in a different 
WB the zeros may not be present  making it difficult to recognise
the ansatz. Furthermore, it was
also shown \cite{Branco:2005jr} that starting from arbitrary 
leptonic mass matrices, the vanishing of certain CP-odd WB invariants,
together with the assumption of no conspiracy among the parameters 
of the Dirac and Majorana mass terms, one is automatically lead 
to given sets of zero textures  in a particular WB.

Frampton, Glashow and Yanagida have shown \cite{Frampton:2002qc}
that it is possible to uniquely relate the sign of the baryon number
of the Universe to CP violation in neutrino oscillation
experiments by imposing two zeros in $m_D$, in the seesaw framework
with only two righthanded neutrino components. 
Two examples were given by these authors:
\begin{eqnarray}
m_D =
\left(
\begin{array}{cc}
a  & 0  \\
a^\prime & b \\
0 & b^\prime
\end{array}\right) \qquad \mbox{or} \qquad
m_D =\left(
\begin{array}{cc}
a  & 0  \\
0 & b \\
a^\prime & b^\prime
\end{array}\right) 
\label{fgy}
\end{eqnarray}
The two zeros in $m_D$ eliminate two CP violating phases, so that
only one CP violating phase remains. This is the most economical
extension of the standard model leading to leptogenesis and
at the same time allowing for low energy CP violation. Imposing that
the model accommodates the experimental facts at low energy
strongly constrains its parameters.

In Ref.~\refcite{Branco:2002xf} minimal scenarios for leptogenesis
and CP violation at low energies were analysed in some specific 
realisations of seesaw models with three righthanded neutrinos and 
four zero textures in $m_D$, where three of the zeros are in the upper
triangular part of the matrix. This latter particular feature was 
motivated by the fact that there is no loss of generality
in parametrising $m_D$ as: 
\begin{equation}
m_D = U\,Y_{\triangle}\,, \label{mDtri}
\end{equation}
with $U$ a unitary matrix and 
$Y_\triangle$ a lower triangular matrix, i.e.:
\begin{equation}
Y_{\triangle}= \left(\begin{array}{ccc}
y_{11} & 0 & 0 \\
y_{21}\,e^{i\,\phi_{21}} & y_{22} & 0 \\
y_{31}\,e^{i\,\phi_{31}} & y_{32}\,e^{i\,\phi_{32}} & y_{33}
\end{array}
\right)\,, \label{Ytri1}
\end{equation}
where $y_{ij}$ are real positive numbers. Choosing  $U=1$
reduces the number of parameters in $m_D$. Moreover, $U$
cancels out in the combination $m^\dagger_D m_D$ relevant in the case
of unflavoured leptogenesis, whilst it does not cancel in $m_{eff}$.
Therefore choosing $U=1$ allows for a connection
between low energy CP violation and leptogenesis to be established
since in this case the same phases affect both phenomena.
The nonzero entries of $m_D$
were written in terms of powers of a small parameter a la Frogatt
Nielsen \cite{Froggatt:1978nt} and chosen in such a way as to 
accommodate the experimental data. Viable leptogenesis
was found requiring the existence of low energy CP violating effects
within the range of sensitivity of the future long baseline neutrino
oscillation experiments under consideration.

In order to understand how the connection between CP violation
required for leptogenesis and low energy physics is established 
in the presence of zeros in the matrix $m_D$, the following
relation derived from Eq.~(\ref{14}) in the WB where $M_R$
and $m_l$ are real positive and diagonal is important:
\begin{equation}
m_D = i U_{\nu} {\sqrt d} R {\sqrt D_R}
\label{udr}
\end{equation}
with $R$ an orthogonal complex matrix,
${\sqrt D_R }$ a diagonal real
matrix verifying the relation
${\sqrt D_R } {\sqrt D_R }= D_R $
and  ${\sqrt d }$ a real matrix with a maximum number
of zeros such that
${\sqrt d} {\sqrt d}^T = d $.
This is the well known Casas and Ibarra parametrisation \cite{Casas:2001sr}.
From this equation it follows that:
\begin{equation}
m^\dagger_D m_D = {\sqrt D_R} R^{\dagger}{\sqrt d}^T {\sqrt d} R
{\sqrt D_R}  \label{drr}
\end{equation}
Since the CP violating phases relevant for leptogenesis in the unflavoured
case are those contained in $m^\dagger_D m_D$, it is
clear that leptogenesis can occur even if there is
no CP violation at low energies i.e. no Majorana- or Dirac- type
CP phases at low energies \cite{Rebelo:2002wj}. Unflavoured leptogenesis
requires the matrix $R$ to be complex. In flavoured
leptogenesis the separate lepton $i$ family asymmetry generated from 
the decay of the $k$th heavy Majorana neutrino  depends on the 
combination \cite{Fujihara:2005pv}
Im$\left( (m_D^\dagger m_D)_{k k^\prime}(m_D^*)_{ik} (m_D)_{ik^\prime}\right) $
as well as on
Im$\left( (m_D^\dagger m_D)_{k^\prime k}(m_D^*)_{ik} 
(m_D)_{ik^\prime}\right) $. The matrix $U_\nu$
does not cancel in each of these terms
and it was shown that it is possible to have viable leptogenesis
even in the case of real $R$, with CP violation in the PMNS matrix
as the source of CP violation required for leptogenesis. 

From Eq.~(\ref{udr}) it is clear that one zero in $(m_{D})_{ij}$ 
corresponds to having an orthogonality relation between 
the ith row of the matrix  $U_{\nu }\sqrt{d}$
and the jth column of the matrix $R$:
\begin{equation}
(m_{D})_{ij}=0~:\qquad (U_{\nu })_{ik}\sqrt{d}_{kl}R_{lj}=0  
\label{orto}
\end{equation}
Ibarra and Ross \cite{Ibarra:2003up} showed that, in the seesaw case 
with only two righthanded neutrinos,
a single zero texture, has the special feature of
fixing the matrix $R$, up to a reflection, without imposing any further
restriction on light neutrino masses and mixing.
The predictions from models with  two 
zero textures in $m_D$ were also analysed in detail in their work, 
including the 
constraints on leptogenesis and lepton flavour violating processes.
The number of all different two texture zeros is fifteen.
Two zeros imply two simultaneous conditions of the type given by
Eq.~(\ref{orto}). Compatibility of these two conditions implies
restrictions on $U_\nu$ and  $\sqrt{m_i}$. Only five of these cases
turned out to be allowed experimentally, including the two cases
of Eq.~(\ref{fgy}) in this reference.

All of these two zero texture ans\"atze satify the following WB 
invariant condition  \cite{Branco:2005jr}:
\begin{equation}
I_1 \equiv \mbox{tr} \left[ m_D M^\dagger_R M_R m^\dagger_D, h_l 
\right] ^3 = 0
\label{inv}
\end{equation}
with  $h_l = m_l m^\dagger_l$, as before.
It was also shown \cite{Branco:2005jr}
that for arbitrary complex leptonic mass matrices,
assuming that there are no special relations among the entries
of $M_R$ and those of $m_D$  this condition automatically leads to one
of the two zero anz\" atze classified in Ref.~\refcite{Ibarra:2003up}. The
assumption that $M_R$ and $m_D$ are not related to each other 
is quite natural, since $m_D$ and $M_R$ originate from different
terms of the Lagrangian.

There are other CP-odd WB invariants which vanish
for all of the two zero textures just mentioned, even if they arise
in a basis where $M_R$ is not diagonal. An example is the
following WB invariant condition  \cite{Branco:2005jr}:
\begin{equation}
I^\prime \equiv \mbox{tr} \left[ m_D m^\dagger_D, h_l 
\right] ^3 =0
\label{wbi}
\end{equation}
which is verified for any texture with two zeros in $m_D$
in a WB where $m_l$ is diagonal, while $M_R$ is arbitrary.

The case of zero textures with three righthanded neutrinos
was also considered in Ref~\refcite{Branco:2005jr}. In this case the
WB invariant $I_1$ always vanishes for three zero textures in $m_D$ with
two orthogonal rows, which implies that one row has no zeros.
The case of three zeros corresponding to two orthogonal columns of
$m_D$, which in this case implies that one column has no zeros
leads to the vanishing of a new invariant $I_2$, defined by:
\begin{equation}
I_{2}\equiv \mathrm{ tr \ }\left[ M_{R}^{\dagger }M_{R}\ ,\ \ m_{D}^{\dagger
}m_{D}\right]^3   \label{i2}
\end{equation}

Four zero textures in the context of seesaw with three righthanded 
neutrinos are studied in detail in Ref.~\refcite{Branco:2007nb}. 
It is shown that 
four is the maximum number of zeros in textures compatible with the
observed leptonic mixing and with the additional requirement that 
none of the neutrino masses vanishes. It is also shown that such 
textures lead to 
important constraints both at low and high energies, and allow 
for a tight connection
between leptogenesis and low energy parameters.
It is possible in all cases to completely
specify the matrix $R$ in terms of light neutrino masses and
the PMNS matrix. These relations are explicitly given in 
Ref.~\refcite{Branco:2007nb}.

\section*{Acknowledgements}
The author thanks the Organizers of the
Sixth International Heidelberg Conference on
Dark Matter in Astro and Particle Physics which took
place in Sydney, Australia for the the warm hospitality
and the stimulating scientifical environment provided.
This work was partially supported by Funda\c c\~ ao para a
Ci\^ encia e a  Tecnologia (FCT, Portugal) through the projects
PDCT/FP/63914/2005, PDCT/FP/63912/2005 and
CFTP-FCT UNIT 777 which are partially funded through POCTI 
(FEDER).

\end{document}